\documentstyle[citesort,epsf,a4,12pt]{article}

\def\lsim{\;\raise0.3ex\hbox{$<$\kern-0.75em\raise-1.1ex\hbox{$\sim$}}\;}
\def\gsim{\;\raise0.3ex\hbox{$>$\kern-0.75em\raise-1.1ex\hbox{$\sim$}}\;}
\newcommand{\ba}{\begin{array}}
\newcommand{\ea}{\end{array}}
\newcommand{\bea}{\begin{eqnarray}}
\newcommand{\eea}{\end{eqnarray}}
\newcommand{\nn}{\nonumber}

\begin{document}

\begin{titlepage}

\begin{flushright}
Orsay LPTHE-98-79\\
hep-ph/9812427
\end{flushright}

\vspace{3cm}
\centerline{\Large\bf Topologies of the (M+1)SSM}
\centerline{\Large\bf with a Singlino LSP at LEP2}

\vspace{2cm}
\begin{center}
{\bf U. Ellwanger and C. Hugonie}\\
Laboratoire de Physique Th\'eorique et Hautes Energies
\footnote{Laboratoire associ\'e au Centre National de la Recherche Scientifique
(URA D0063)}\\    
Universit\'e de Paris XI, Centre d'Orsay, F-91405 Orsay Cedex, France\\  
\end{center}
\vspace{2cm}

\begin{abstract}

We study the possible signals of the (M+1)SSM with a singlino LSP at LEP2.
First we identify regions of the parameter space which are ruled out by
negative results of sparticle searches in the context of the MSSM. In the
remaining kinematically accessible regions we present total event rates for
topologies which require further studies, i.e. estimations of the corresponding
efficiencies: various 4 charged fermion final states with missing energy,
possibly with displaced vertices due to a long lifetime of the NLSP, the second
lightest neutralino. Searches for these unconventional signatures are essential
in order to cover the entire kinematically accessible parameter space of the
(M+1)SSM with a singlino LSP at LEP2. 

\end{abstract}

\end{titlepage}

\section{Introduction}

The supersymmetric extension of the standard model with an additional gauge
singlet superfield, the so called (M+1)SSM
\cite{UMC,NMSSM1,NMSSM2,Higgs,walls,neu2,RGE1,Steph,last}, solves naturally the
$\mu$-problem of the MSSM: Even for a scale invariant superpotential -- with a
coupling $\lambda S H_1 H_2$ among the Higgs superfields and the singlet
superfield $S$ -- an effective $\mu$-term $\mu = \lambda \langle S \rangle$ is
generated, if the scalar component of $S$ has a non vanishing vev. Such a vev
of the order of the weak scale can be generated through the standard soft
supersymmetry breaking terms, thus the weak scale appears exclusively in the
form of the supersymmetry breaking scale. Moreover, assuming universal soft
terms at a large (GUT) scale, the (M+1)SSM has the same number of free
parameters as the MSSM. Previous analyses of the parameter space of the model
\cite{NMSSM2,UMC,last} have shown that, as in the case of the MSSM, a large
region is consistent with the present experimental bounds on sparticle and
Higgs masses. 

The particle content of the (M+1)SSM differs from the MSSM in the form of
additional gauge singlet states in the Higgs sector (1 neutral CP-even and 1
CP-odd state) and in the neutralino sector (a two component Weyl fermion).
These states mix with the corresponding ones of the MSSM, with a mixing angle
which is proportional to the coupling $\lambda$ above. Accordingly, the
phenomenology of the (M+1)SSM depends on a large extend on the magnitude of
$\lambda$: 

For $\lambda \gsim O(10^{-2})$ the masses and couplings, notably in the CP-even
Higgs sector, can deviate visibly from the ones of the MSSM \cite{Higgs};
however, in this region of the parameter space of the (M+1)SSM some fine-tuning
among the parameters is required in order to meet all the phenomenological
constraints \cite{UMC}. 

For $\lambda \lsim O(10^{-2})$ the mixing angles involving the singlet states
are quite small. Therefore, the Higgs and sparticle masses and couplings of the
(M+1)SSM are very close to the MSSM ones (for corresponding values of $\mu$ and
$B$ \cite{NMSSM2,UMC}), with additional quasi singlet states which have small
couplings to the gauge bosons and the MSSM sparticles. Accordingly, they have
small production cross sections, and they will not appear in sparticle decays
unless they represent the only kinematically allowed decay channel. 

Assuming R parity conservation, this latter situation is realized if the quasi
singlet Weyl fermion (the singlino) is the LSP. Then the singlino will appear
in the final state of each sparticle decay, and the phenomenology of the
(M+1)SSM with a singlino LSP differs considerably from the one of the MSSM. 

In a previous paper \cite{last} we have shown that this situation appears
naturally in the case of a gaugino dominated supersymmetry breaking: $M_{1/2}
\gg A_0,m_0$. Then, within the parameter space accessible at LEP2, the NLSP is
mostly a bino-like state. Hence all the processes involving sparticle
productions and decays will end up with a bino to singlino transition, and we
have studied the corresponding decay widths in \cite{last}. An important result
was that, for small values of $\lambda$ or for singlino masses close to the
bino mass, the bino life time can be so large that the bino to singlino cascade
appears at macroscopic distances from the production point, or even out of the
detector. 

In the present paper we study the possible signals of the (M+1)SSM with a
singlino LSP at LEP2 in the various regions of the parameter space. First we
identify those regions which are ruled out by negative results of sparticle
searches in the context of the MSSM. In the remaining kinematically allowed
regions we present total event rates for various topologies, like 4 charged
fermion final states and missing energy, with or without displaced vertices.
Such topologies, with microscopic vertices, have been looked for at LEP2 in the
context of the MSSM or models with R parity violation. However the
corresponding efficiencies do not apply to the (M+1)SSM with a singlino LSP.
With estimated efficiencies, we find that considerable kinematically allowed
regions of the parameter space have not been tested at present, especially in
the case of macroscopically displaced vertices. The main purpose of the present
paper is to identify those topologies, for which further studies -- i.e.
estimation of efficiencies -- are required in order to interpret the available
or incoming data from LEP2 in the context of the (M+1)SSM with a singlino LSP. 

It is a priori not clear whether negative results of sparticle searches would
constrain the (M+1)SSM with a singlino LSP more or less than the MSSM: The
final states associated with the pair production of a given sparticle (like the
selectron or chargino) will often be more involved in the (M+1)SSM as compared
to the MSSM, and the corresponding constraints on the cross sections are often
much weaker. On the other hand, the (M+1)SSM with a singlino LSP allows for a
process to be observable, which is invisible within the MSSM: the production of
a pair of binos. If the binos decay into singlinos plus additional observable
particles, LEP2 is sensitive to light binos, which would, however, escape
detection within the associated MSSM. (Here and below the associated MSSM
denotes the MSSM obtained after "freezing" the singlet vev, which generates
effective $\mu$ and B terms, and after dropping the gauge singlet states in 
the neutralino and Higgs sectors.)  
Thus the application of the LEP2 results to the (M+1)SSM
requires a case by case analysis, depending on the different regions of the
parameter space, which will be performed below. 

In order to scan the complete parameter space of the (M+1)SSM we proceed as in
\cite{UMC,last}: First we assume universal scalar masses $m_0$, gaugino masses
$M_{1/2}$ and trilinear couplings $A_0$ at the GUT scale. Thus we scan over the
ratios $m_0/M_{1/2}$, $A_0/M_{1/2}$ and the Yukawa couplings at the GUT scale,
the absolute scale being determined at the end by requiring the correct value
of $M_Z$. For each point in the parameter space we integrate the
renormalization group equations \cite{RGE1} down to the weak scale, and
minimize the low energy effective potential including the full one loop
radiative corrections \cite{Higgs}. We check whether squarks or sleptons do not
assume vevs, diagonalize numerically the mass matrices and verify whether
applicable bounds on sparticle and Higgs masses are satisfied. 

In contrast to \cite{UMC,last}, however, we have included as matter Yukawa
couplings not just the top Yukawa coupling $h_t$, but all the couplings of the
third generation $h_t$, $h_b$ and $h_\tau$. First, this makes our results more
reliable in the large $\tan(\beta)$ regime, and second this reveals a new
phenomenon: Within the (M+1)SSM with a singlino LSP and sparticle masses in the
reach of LEP2, the NLSP could possibly be the lightest stau $\widetilde\tau_1$.
(In the associated MSSM the lightest stau $\widetilde\tau_1$ would then be the
true LSP, i.e. a stable charged particle; this situation has been discussed in 
\cite {mura}.)

The paper is organized as follows: In the next section we present the
lagrangian and discuss the different regions in the parameter space which are
relevant for the present investigations. In section three we study the
sparticle production processes which are kinematically allowed at LEP2, the
topologies relevant for searches in the context of the (M+1)SSM with a singlino
LSP, and the constraints on its parameters which could be already infered from
available data. The total number of events expected in those regions of
parameter space is given, for which the efficiencies still remain to be
determined. Conclusions are presented in section four.

\section{Parameter space of the (M+1)SSM with a singlino LSP} \label{secparam} 

The superpotential of the (M+1)SSM is given by
\bea
W & = & \lambda SH_1H_2 + \frac{1}{3}\kappa S^3 + h_t Q_3H_1U_{3R}^c \nn \\ 
& & + h_b Q_3H_2D_{3R}^c + h_\tau L_3H_2E_{3R}^c + \ldots \label{spot}
\eea
where $Q_3$ denotes the left handed doublet of quarks of the third generation,
$U_{3R}^c$ and $D_{3R}^c$ the (charge conjugate) right handed top and bottom
quarks, $L_3$ the left handed doublet of leptons of the third generation,
$E_{3R}^c$ the (charge conjugate) right handed tau. The ellipses in
(\ref{spot}) denote Yukawa couplings involving quarks and leptons of the first
two generations. The only dimensionful parameters of the model are the
supersymmetry breaking parameters (for simplicity, we do not display the terms
involving squarks and sleptons): 
\bea
{\cal L}_{soft} & = & \frac{1}{2} \left( M_3\lambda_3^a\lambda_3^a +
M_2\lambda_2^i\lambda_2^i + M_1\lambda_1\lambda_1 \right) + \mbox{h.c.} \nn
\\ 
& & - m_1^2|H_1|^2 - m_2^2|H_2|^2 - m_S^2|S|^2 \nn \\
& & - \lambda A_\lambda SH_1H_2 - \frac{1}{3}\kappa A_\kappa S^3 + \mbox{h.c.}
\label{Lsoft} 
\eea
where $\lambda_3$, $\lambda_2$ and $\lambda_1$ (the 'bino') are the gauginos of
the $SU(3)_c$, $SU(2)_L$ and $U(1)_Y$ gauge groups respectively. The scalar
components of the Higgs in (\ref{Lsoft}) are denoted by the same letters as the
corresponding chiral superfields. These supersymmetry breaking terms are
constrained in the present version of the model by universality at the scale
$M_{GUT} \sim 10^{16}$~GeV. Thus, the independent parameters are: Universal
gaugino masses $M_{1/2}$ (always positive in our convention); universal masses
$m_0^2$ for the scalars; universal trilinear couplings $A_0$ (either positive
or negative); the Yukawa couplings $h_{t0}$, $h_{b0}$, $h_{\tau 0}$,
$\lambda_0$, $\kappa_0$ of the superpotential (\ref{spot}) at the scale
$M_{GUT}$. 

The parameters at the weak scale are obtained by integrating numerically the
one loop renormalization group equations \cite{RGE1}. The Coleman-Weinberg
radiative corrections to the effective potential involving top/stop,
bottom/sbottom and tau/stau loops (beyond the leading log approximation)
\cite{Higgs} are taken into account. The results for the mass matrices, after
minimization of the effective potential, can be found in
\cite{NMSSM1,RGE1,Higgs,NMSSM2,UMC,last} and will not be repeated here. Mixing
terms are considered in the stop, sbottom and stau mass matrices. 

Let us now discuss the parameter space of the (M+1)SSM with a singlino LSP
which is relevant for sparticle searches at LEP2. Since here the Yukawa
couplings $\lambda$ and $\kappa$ are quite small ($\lambda , \kappa \lsim
O(10^{-2})$) and hence the singlet sector mixes only weakly to the non singlet
sector, it is possible to understand the gross features of the parameter space
with the help of analytic approximations to the integrated renormalization
group equations, the minimization of the effective potential and the mass
matrices \cite{RGE2,NMSSM2,UMC,last}. (The results in section \ref{sectop}, on
the other hand, are based on 'exact' numerical computations for $\sim 10^4$
points in the parameter space.) 

First, we consider the neutralino sector. In our convention, the (symmetric)
neutralino mass matrix is given by \cite{HK} 
\bea
M^0 = \left( \ba{ccccc}
	M_2 & 0 & \displaystyle{\frac{-g_2h_1}{\sqrt{2}}} &
\displaystyle{\frac{g_2h_2}{\sqrt{2}}} & 0 \\ 
	& M_1 & \displaystyle{\frac{g_1h_1}{\sqrt{2}}} &
\displaystyle{\frac{-g_1h_2}{\sqrt{2}}} & 0 \\ 
	& & 0 & -\mu & -\lambda h_2 \\
	& & & 0 & -\lambda h_1 \\
	& & & & 2\kappa s \ea \right) . \label{masneu}
\eea

For small $\lambda$, the singlino is thus an almost pure state of mass 
\bea
M_{\widetilde S} \simeq 2 \kappa s .
\eea
and the vev $s$ of the scalar singlet can be estimated from the tree level
scalar potential: 
\bea
s \simeq -\frac{A_\kappa}{4\kappa} \left( 1 + \sqrt{1-\frac{8m_S^2}
{A_\kappa^2}} \right) . \label{svev} 
\eea

Since $A_\kappa$ and $m_S$ are only slightly renormalized between $M_{GUT}$ and
the weak scale for small $\lambda$ and $\kappa$, $M_{\widetilde S}$ can be
written in terms of the universal soft terms at $M_{GUT}$: 
\bea
M_{\widetilde S} \simeq -\frac{A_0}{2}\left( 1 + \sqrt{1-\frac{8m_0^2}{A_0^2}}
\right) . \label{msing} 
\eea

The condition for the minimum (\ref{svev}) to be deeper than the trivial one
reads at tree level
\bea
|A_0| > 3m_0 \label{A0m0}
\eea
so that
\bea
\frac{2}{3}|A_0| \lsim |M_{\widetilde S}| \lsim |A_0| .
\eea

Since the effective $\mu$ parameter turns out to be quite large, 
\bea
\mu^2 = \lambda^2 s^2 \simeq 2.5 M_{1/2}^2 -.5 M_Z^2 \label{mu},
\eea
the lightest non singlet neutralino is the (nearly pure) bino $\widetilde B$
with mass $M_{\widetilde B}$. From the approximate analytic diagonalization of
(\ref{masneu}) for $\tan(\beta) \gsim 5$ (which, from our numerical results, is
always the case for a singlino LSP), one obtains $M_{\widetilde B}$ in terms of
the universal gaugino mass $M_{1/2}$ as 
\bea
M_{\widetilde B} & \simeq & M_1 + \frac{\sin^2\theta_WM_Z^2M_1}{M_1^2-\mu^2} \nn
\\ 
& \simeq & .41M_{1/2} - \frac{4.10^{-2}M_Z^2M_{1/2}}{M_{1/2}^2-.2M_Z^2}
\label{mbino} 
\eea
where we have used (\ref{mu}) and $M_1 = .41 M_{1/2}$. The second term in
(\ref{mbino}) is due to the bino/higgsino mixing. From (\ref{msing}) and
(\ref{mbino}) one finds that the necessary (resp. sufficient) conditions on the
universal terms for a singlino LSP are 
\bea
|A_0| \lsim .6 M_{1/2} \quad ( \mbox{resp. } |A_0| \lsim .4 M_{1/2} ) .
\label{A0M1/2} 
\eea

The Yukawa couplings $\lambda$ and $\kappa$ of the (M+1)SSM are, in general,
constrained by the ratio $A_0/M_{1/2}$. From the absence of a deeper unphysical
minimum of the Higgs potential with $h_2=s=0$ the following inequality can be
derived: 
\bea
\kappa \lsim 4.10^{-2}\frac{A_0^2}{M_{1/2}^2} .
\eea

Since the singlet vev $s$ increases with decreasing $\kappa$ (cf.
(\ref{svev})), but the effective $\mu$ term should be of the order of the weak
scale, $\lambda$ and $\kappa$ should be of the same order of magnitude. From
our numerical analysis we find that the bare parameters $A_0$, $M_{1/2}$ and
$\lambda_0$ satisfy the (not very stringent) relation 
\bea
\frac{|A_0|}{M_{1/2}} \sim 4 \lambda_0^{.5\pm.3} ;
\eea
thus light singlinos are generally related to small values of $\lambda$ and
$\kappa$. Since the mixing angle of the singlino to the non singlet sector is
proportional to $\lambda$, all decay widths of sparticles into a singlino LSP
are at least proportional to $\lambda^2$. Furthermore, $\lambda$ can be
extremely small; then the NLSP life time is very large. This phenomenon,
already investigated in \cite{last}, will play an important role in the next
section. 

Now we turn to the slepton sector. The lightest states are the 'right handed'
charged sleptons $\widetilde l_R$ and the sneutrinos $\widetilde\nu$. Since the
bare scalar mass $m_0$ is quite small (cf. (\ref{A0m0}) and (\ref{A0M1/2})) the
corresponding mass terms at the weak scale are determined, from the integrated
renormalization group equations, by $M_{1/2}$. Neglecting the mixing between
the right handed and the left handed sleptons, and using the known numerical
values of the electroweak gauge couplings appearing in the D terms, their
masses are (for medium or large $\tan(\beta)$) 
\bea
m_{\widetilde l_R}^2 = & m_E^2 - \sin^2\theta_W M_Z^2 \cos 2\beta & \simeq .15
M_0^2 + .23 M_Z^2 , \label{msel} \\ 
m_{\widetilde\nu}^2 = & m_L^2 + \frac{1}{2}M_Z^2 \cos 2\beta & \simeq .52 M_0^2
- .5 M_Z^2 . \label{msneu} 
\eea

The limit on the sneutrino mass obtained from the $Z$ width, $m_{\widetilde\nu}
\gsim M_Z/2$ \cite{Griv}, combined with (\ref{msneu}) gives a lower limit on
$M_{1/2}$: 
\bea
M_{1/2} \gsim 100\mbox{ GeV} . \label{M1/2min}
\eea

From (\ref{msel}) together with (\ref{mbino}) it follows that the sleptons
$\widetilde l_R$ are heavier than the bino for $M_{1/2} \lsim 320$~GeV.
However, this result holds only for the charged sleptons of the first two
generations. For the third generation, the soft masses at low energy can be
smaller than the ones given in (\ref{msel}) and (\ref{msneu}) (depending on
$h_\tau$). Furthermore, the off-diagonal term in the stau mass matrix is given
by $h_\tau(\mu h_1 - A_\tau h_2)$, which is not necessarily negligible compared
to the smallest diagonal term. Thus, the lightest eigenstate $\widetilde\tau_1$
will be lighter than the right handed sleptons of the first two generations
$\widetilde l_R$ and can well be lighter than the bino even for $M_{1/2} \lsim
320$~GeV (hence for sparticle masses within the reach of LEP2). 

In the chargino sector, within the present region of the parameter space, the
lightest eigenstate is essentially a wino of mass $M_2$ given in terms of
$M_{1/2}$ by 
\bea
M_2 = .82 M_{1/2} . \label{mcharg}
\eea

In the Higgs sector we can again make use of the fact that the non singlet and
singlet sectors are quasi decoupled. The direct search for Higgs scalars thus
proceeds as in the MSSM, and the present negative results do not impose more
stringent constraints on $M_{1/2}$ than (\ref{M1/2min}). (For large values of
$\lambda$, without singlino LSP, the Higgs phenomenology of the (M+1)SSM could,
however, differ substantially from the one of the MSSM \cite{Higgs}.) 

Since the scalar Higgs quasi singlet state can possibly be produced in bino
decays in the (M+1)SSM, its mass $M_S$ will be of interest. From the tree level
part of the Higgs potential one finds for small Yukawa couplings 
\bea
M_S^2 \simeq \frac{1}{4} \sqrt{A_0^2-8m_0^2} \left( |A_0| + \sqrt{A_0^2-8m_0^2}
\right) , 
\eea
hence
\bea
M_S \lsim \frac{|A_0|}{\sqrt 2} .
\eea

For later use we note that the coupling Higgs singlet - bino - singlino is
proportional to $\lambda^2$, thus the production of the Higgs singlet state in
bino decays will only occur for $\lambda$ not too small. 

To summarize, the parameter space of the (M+1)SSM with a singlino LSP is
characterized by the universal gaugino mass $M_{1/2}$ being the dominant soft
supersymmetry breaking term. Both $A_0$ and, consequently, $m_0$ are bounded
from above in terms of $M_{1/2}$ by (\ref{A0M1/2}) and (\ref{A0m0}),
respectively. The Yukawa couplings $\kappa$ and $\lambda$ also have upper
limits of $O(10^{-2})$, and are possibly tiny. 

The non singlet sparticles (with sizeable production cross sections) within the
reach of LEP2 are: The second lightest neutralino, essentially a bino
$\widetilde B$; the right handed sleptons $\widetilde l_R$ with masses given by
(\ref{msel}) and the lightest stau $\widetilde\tau_1$ which could be
substantially lighter; sneutrinos with masses given by (\ref{msneu}); and the
lightest chargino with a mass given by (\ref{mcharg}). Note that, for a value
of $M_{1/2}$ corresponding to a bino in the reach of LEP2, the bino is always
lighter than these sparticles, with the possible exception of the lightest stau
$\widetilde\tau_1$. 

In the next section, we will discuss the different decays of these particles,
and compare the respective final states to sparticle searches at LEP2. This
will allow us to find out which parameter ranges of the (M+1)SSM have been
already ruled out, and which require further study.

\section{Topologies for sparticle searches at LEP2} \label{sectop}

\subsection{Bino decays with a singlino LSP} \label{3.1}

Sparticle searches in the (M+1)SSM with a singlino LSP differ in several
respects from sparticle searches in the MSSM: First, the presence of the
singlino LSP usually gives rise to additional cascades in sparticle decays. For
instance, pair production of binos is usually an observable process, whereas
for an equivalent MSSM (with comparable soft supersymmetry breaking terms), the
bino would correspond to the LSP, and this process would be invisible. Thus,
areas in the soft SUSY breaking parameter space accessible at LEP2 are larger
in the (M+1)SSM than in the MSSM, provided an adapted experimental analysis is
done. Second, the decay of the NLSP (the bino or the lightest stau) into the
singlino LSP is always proportional to a power of $\lambda$, which can be tiny.
In this case (or if the singlino LSP happens to be close in mass to the NLSP,
which is feasible in the (M+1)SSM with universal soft terms in contrast to the
MSSM) the NLSP to LSP transition can be rather slow, leading to macroscopically
displaced vertices. 

In the following we can make use of the fact that the masses of most sparticles
in the (M+1)SSM with a singlino LSP depend essentially on just one parameter,
the universal gaugino mass $M_{1/2}$: For $M_{1/2}$ not too large ($M_{1/2}
\lsim 180$~GeV) $\widetilde B$, $\widetilde l_R$, $\widetilde\nu$ and
$\widetilde\chi_1^\pm$ can be light enough for pair production being
kinematically allowed at LEP2, cf. the dependence of their mass on $M_{1/2}$ in
section \ref{secparam}. On the other hand, for 180~GeV$\lsim M_{1/2} \lsim
220$~GeV, only $\widetilde B$ pair production is kinematically feasible (with
the possible exception of staus). 

Since all the sparticle decays in the (M+1)SSM with a singlino LSP proceed via
the decay of the bino $\widetilde B$ into the singlino $\widetilde S$ (with the
possible exception of the stau $\widetilde\tau_1$, see below), we will briefly
discuss the possible final states of this transition, using the results of
\cite{last}: 

a) $\widetilde B\to \widetilde S \nu\bar\nu$: This invisible process is
mediated dominantly by sneutrino exchange. Since the sneutrino mass, as the
mass of $\widetilde B$, is essentially fixed by $M_{1/2}$ (cf. (\ref{msneu})),
the associated branching ratio varies in a predictable way with $M_{\widetilde
B}$: It can become up to 90\% for $M_{\widetilde B} \sim 30$~GeV, but decreases
with $M_{\widetilde B}$ and is maximally 10\% for $M_{\widetilde B} \gsim
65$~GeV. 

b) $\widetilde B \to \widetilde S l^+l^-$: This process is mediated dominantly
by the exchange of a charged slepton in the s-channel. If the lightest stau
$\widetilde\tau_1$ is considerably lighter than the sleptons of the first two
generations, the percentage of taus among the charged leptons can well exceed
$\frac{1}{3}$. If $\widetilde\tau_1$ is lighter than $\widetilde B$, it is
produced on-shell, and the process becomes $\widetilde B \to \widetilde\tau_1
\tau \to \widetilde S \tau^+ \tau^-$. Hence we can have up to 100\% taus among
the charged leptons and the branching ratio of this channel can become up to
100\%. 

c) $\widetilde B\to \widetilde S S$: This two-body decay is kinematically
allowed if both $\widetilde S$ and $S$ are sufficiently light. (A light $S$ is
not excluded by Higgs searches at LEP1 \cite{LEP1h,LEP2h}, if its coupling to
the $Z$ is too small \cite{Higgs}). However, the coupling $\widetilde B
\widetilde S S$ is proportional to $\lambda^2$, whereas the couplings appearing
in the decays a) and b) are only of $O(\lambda)$. Thus this decay can only be
important for $\lambda$ not too small. In \cite{last}, we found that its
branching ratio can become up to 100\% in a window $10^{-3} \lsim \lambda \lsim
10^{-2}$. Hence,  its length of flight is never macroscopic. Of course, $S$
will decay immediately  into $b\bar b$ or $\tau^+ \tau^-$, depending on its
mass. (If the branching  ratio $Br(\widetilde B\to \widetilde S S)$ is
substantial, $S$ is never lighter than  $\sim 5$~GeV.) If the singlet is heavy
enough, its $b\bar b$  decay gives rise to 2 jets with $B$ mesons, which are
easily detected with $b$-tagging. (However, if the singlet mass is just above
the $b\bar b$ threshold -- typically, if $m_\Upsilon < M_S \lsim 15$~GeV -- $S$
could decay  hadronically without $B$ mesons.) In any case, the hadronic system
-- or the  $\tau^+ \tau^-$ system -- would have an invariant mass peaked at
$M_S$, making this signature easy to search for.

d) $\widetilde B\to \widetilde S \gamma$: This branching ratio can be important
if the mass difference $\Delta M \equiv M_{\widetilde B} - M_{\widetilde S}$ is
small ($\lsim 5$~GeV). 

Further possible final states like $\widetilde B \to \widetilde S q\bar q$ via
$Z$ exchange have always branching ratios below 10\% and will not be considered
here. 

\subsection{Constraints from MSSM-like selectron searches} \label{3.2}

Let us first consider the region in the parameter space where the invisible
decay a) of $\widetilde B$ dominates, which occurs for $M_{1/2} \lsim 140$~GeV.
Then, right handed selectrons $\widetilde e_R$ are light enough for being pair
produced, and they decay as in the MSSM into an electron and a bino, which is
invisible regardless of its lifetime. Results of searches for selectrons with
MSSM-like decays have been published by Aleph \cite{2eA}, Delphi \cite{2eD}, L3
\cite{2eL} and Opal \cite{2eO}\footnote{In this paper, we use the results from
the LEP2 run at $\sqrt{s} = 181-184$~GeV. For recent updates at $\sqrt{s} =
189$~GeV, see Refs.~\cite{LEP189}}. Here, however, the analysis of the results
differs from the situation in the MSSM in two respects: 

First, for a given mass of the selectron, the mass difference $m_{\widetilde
e_R}-M_{\widetilde B}$ is essentially known: for $m_{\widetilde e_R} = 65$~GeV,
e.g., we have $m_{\widetilde e_R}-M_{\widetilde B} \sim 20-30$~GeV. It turns
out that for the mass differences given in the present model, the experimental
efficiencies are always $\gsim 50\%$. 

On the other hand, the branching ratio associated with the invisible decay of
$\widetilde B$ is never 100\%. (Even for $M_{1/2} \lsim 140$~GeV, $\widetilde
B$ could still decay dominantly into $\widetilde S S$, if $\lambda$ happens to
be in the window $10^{-3} \lsim \lambda \lsim 10^{-2}$.) Thus, for each point
in the parameter space, we have to calculate the expected number of MSSM-like
events (2 electrons and missing energy) taking the corresponding branching
ratio into account. 

The most detailed informations on the efficiencies, the numbers of background
and observed events, as a function of $m_{\widetilde e_R}$ and $M_{\widetilde
B}$, are given by Opal \cite{2eO}. From these results, we find that points in
the parameter space leading to $N \gsim 10$ expected events with 2 acoplanar
electrons in the final state are excluded. This occurs in the region 
\bea
M_{1/2} \lsim \mbox{125 GeV or } M_{\widetilde B} \lsim \mbox{43 GeV} .
\label{liminv} 
\eea
However, this region is not totally excluded by acoplanar electron searches: As
mentioned above, $\widetilde B$ could still decay dominantly into $\widetilde S
S$, if $\lambda$ happens to be in the window $10^{-3} \lsim \lambda \lsim
10^{-2}$. 

Further MSSM-like processes associated with 2 leptons and missing energy in the
final state do not lead to additional constraints on the parameter space. 

\subsection{Higher multiplicity final states without displaced vertices}
\label{3.3}

Next, we have to take into account visible $\widetilde B$ cascade decays,
leading to events with higher multiplicity.  First we treat the case where all
sparticle decays take place within at most 1~ cm around the primary vertex,
i.e. $\lambda$ and $\Delta M $ not too small. The following pair production
processes have to be considered: 
\bea
\ba{llclclcl}
\mbox{p.1:} & e^+ e^- & \to & \widetilde B \widetilde B , \\
\mbox{p.2:} & e^+ e^- & \to & \widetilde l_R \widetilde l_R^* & \to & l^+
\widetilde B l^- \widetilde B , \\ 
\mbox{p.3:} & e^+ e^- & \to & \widetilde\nu \widetilde\nu^* & \to & \nu
\widetilde B \bar\nu \widetilde B , \\ 
\mbox{p.4:} & e^+ e^- & \to & \chi_1^+ \chi_1^- & \to & l^+ \widetilde\nu l'^-
\widetilde\nu^* & \to & l^+ \nu \widetilde B l'^- \bar\nu \widetilde B .
\ea \label{proc}
\eea

Taking the bino decays a) -- c) in sect. 3.1 into account, the possible final
states are those listed in Table 1. (The radiative decay $\widetilde B \to
\widetilde S  \gamma$ will be discussed below.)

Let us first consider processes with 4 visible fermions and missing energy. The
appropriate cascade decays of the binos leading to 4 charged fermions in the
final state are: visible decays b) $\widetilde B \to \widetilde S l^+ l^-$ or
c) $\widetilde B \to \widetilde S S \to \widetilde S b \bar b \mbox { or }
\widetilde S \tau^+\tau^-$ for the 2 binos in p.1 and p.3 (the final states
(i.1) and (i.3), i = 5 \dots 9, in Table 1); one bino decaying invisibly
through channel a) $\widetilde B\to \widetilde S \nu\bar\nu$, the other
decaying into $l^+ l^-$ or $b \bar b$ and missing energy through channels b) or
c) for p.2 and p.4 (the final states (i.2) and (i.4), i = 2,3,4, in Table 1).
In the case of the process p.4 we have used the fact  that sneutrinos are
always lighter than the lightest chargino in the (M+1)SSM with a singlino LSP,
thus the latter decays exclusively into an on-shell sneutrino and a charged
lepton. 

According to the discussion of the decay channel b) above, the charged leptons
$l^\pm$ in the final state can be the leptons of any generation. In the case of
light staus, the percentage of taus among the charged leptons can become up to
100\%. If the lightest stau $\widetilde\tau_1$ is the NLSP, p.2 and p.4 give 6
charged leptons plus missing energy in the final state. Only p.1 and p.3 lead
to 4 charged leptons (taus) plus missing energy, since, in this case, the only
decay channel for the bino is $\widetilde B \to \tau \widetilde\tau_1 \to
\widetilde S \tau^+ \tau^-$. 

Thus, the final states of interest are $l^+ l^- l^+ l^-$, $l^+ l^- b \bar b$
and $b \bar b b \bar b$ plus missing energy. Since the $b$s can arise solely
from the decay c) $\widetilde B \to \widetilde S S \to \widetilde S b \bar b$,
the invariant mass of a $b \bar b$ system would always be peaked at $M_S$, cf.
the discussion above. However, for a given value of $M_{1/2}$, we cannot
predict the different branching ratios of $\widetilde B$ ($\lambda$ may or may
not be in the window where the decay into $\widetilde S S$ is dominant), hence
we cannot predict the ratios of the different final states associated to a
given process in (\ref{proc}). On the other hand, for a given value of
$M_{1/2}$ we know, with small errors, the masses $M_{\widetilde B}$,
$m_{\widetilde l_R}$, $m_{\widetilde\nu}$ and $M_{\chi_1^\pm}$ and the
corresponding production cross sections. For each point in the parameter space
obtained from the scanning described in the previous section, we have
calculated numerically the production cross sections of the proceses p.1-4,
taking into account possible interference terms between s-, t- and u-channels
\cite{cross}, for  $e^+ e^-$ collisions at 183~GeV c.m. energy. In
Fig.~\ref{fig1} we show, for each point in the parameter space, the total
number of events with 4 charged fermions plus missing energy in the final state
as a function of $M_{1/2}$, assuming an integrated luminosity  of 55~pb$^{-1}$.
We have already removed those points in the parameter space where $\widetilde
B$ decays dominantly invisibly through channel a), and which are excluded by
the negative results of selectron searches in the MSSM, see the discussion
above. Moreover, we have not shown the points where $\widetilde B$ decays
dominantly into channel d) $\widetilde B \to \widetilde S \gamma$ which will be
discussed separately below. 

In Fig.~\ref{fig1} we observe a large number of events for $M_{1/2} \lsim
150$~GeV, which are due to the process p.3: If kinematically allowed, its cross
section is typically larger than the ones of p.1, p.2 or p.4. For $M_{1/2}
\gsim 150$~GeV, on the other hand, the number of events is essentially given by
the number of $\widetilde B$ being pair produced (p.1). 

Events with 4 charged fermions plus missing energy in the final state have been
searched for at LEP2. The underlying processes were assumed to be: $\widetilde
t_1$ pair production with $\widetilde t_1 \to b l \widetilde\nu$
\cite{stop3b,2eD} and heavy neutralinos decaying via the Multi Lepton channel
\cite{neuML} in the MSSM; lightest neutralino pair production in models with
gauge mediated supersymmetry breaking (i.e. a gravitino LSP) and a stau NLSP
\cite{GMSBstau}; or any sparticle pair production process in the context of
models with R parity violation \cite{Rp}.

Standard backgrounds with 4 charged fermions and missing energy are small and
typically, after imposing appropriate cuts, the number of background events in
a given channel vary from 0 to 4, with a comparable number of observed events.
No excess has been observed. The given efficiencies vary roughly between 20\%
and 60\% depending, e.g. in ${/ \hskip - 3 truemm R_p}$ models, on the mass of
the unstable (intermediate) neutralino. 

Of course we cannot apply these efficiencies to the processes listed in
(\ref{proc}). The kinematics of these processes is often very different from
the kinematics of the assumed underlying processes, and also various branching
ratios into different final states would have to be considered. (In particular
in the case of small mass difference $\Delta M$ the efficiencies for the
processes p.1-p.4 could be quite low.) 

From Fig.~\ref{fig1} we can only deduce which range of values for $M_{1/2}$
could be excluded. For instance, assuming a minimal efficiency of 20\% for all
processes listed in (\ref{proc}), and assuming a total number of 4 expected
events excluded, we would conclude that the total number of actual events has
to be smaller than 20 implying a lower limit on $M_{1/2}$ or $M_{\widetilde B}$
of 
\bea
M_{1/2} \gsim \mbox{190 GeV or } M_{\widetilde B} \gsim \mbox{75 GeV} .
\eea
(In Fig.~\ref{fig1} we have indicated this example by a horizontal line.)

Events with 6 charged fermions in the final state can also appear in slepton or
chargino pair production (processes p.2 and p.4, the final states (i.2) and
(i.4), i = 5 \dots 9, in Table 1). However, the bino is always lighter than
these sparticles (with the possible exception of the stau, see below), and the
regime in the parameter space covered by $\widetilde B$ pair production (and 4
charged fermions in the final state) is always larger.

Next, we comment briefly the case d) where $\widetilde B$ decays dominantly
into $\widetilde S \gamma$. Note that this branching ratio can only be
important for a small mass difference $\Delta M = M_{\widetilde B} -
M_{\widetilde S} \lsim 5$~GeV. This decay could lead to final states with just
2 isolated photons and missing energy (via p.1 and p.3) or 2 leptons plus 2
isolated photons and missing energy (via p.2 and p.4). In the first case ,
however, detection efficiencies are always very small due to the small mass
difference $\Delta M$ \cite{2gA,2gD,2gL,2gO}. Final states of the form $l^+ l^-
\gamma \gamma + {/ \hskip - 3 truemm E_T}$ have been searched for in
\cite{2eA,GMSB,chargg}, where gauge mediated supersymmetry breaking (i.e. a
gravitino LSP) was assumed. Again, however, the efficiencies corresponding to
the assumed underlying process do not apply to the present case due to the
small value of $\Delta M$. On the other hand, if the photons are soft enough to
be accepted as low energy neutral clusters in acoplanar lepton searches, the
MSSM constraint on the selectron mass $m_{\widetilde e_R} \lsim 80$~GeV
\cite{2eA,2eD,2eL,2eO} applies, leading to a lower limit on $M_{1/2}$
($M_{\widetilde B}$) of 
\bea
M_{1/2} \gsim \mbox{175 GeV or } M_{\widetilde B} \gsim \mbox{67 GeV}
\label{limsel}. 
\eea
Clearly this case requires a dedicated analysis depending on the various
detectors. 

\subsection{Final states with neutral displaced vertices} \label{3.4}

Up to now, we have considered the case of a microscopic lifetime of $\widetilde
B$. For a small Yukawa coupling $\lambda$ or a small $\Delta M$, however, the
length of flight of $\widetilde B$ can become large, leading to macroscopically
displaced vertices \cite{last}. Let us first remark that, in this case, the
decay channel c) $\widetilde B \to \widetilde S S$ is impossible: If the decay
length of $\widetilde B$ is large, either $\lambda$ is very small and thus
outside the window $10^{-3} \lsim \lambda \lsim 10^{-2}$, or $\Delta M$ is
small such that the quasi-singlet Higgs scalar $S$ can no longer be produced on
shell. Furthermore, the region of the parameter space where the invisible decay
channel a) $\widetilde B \to \widetilde S \nu \bar\nu$ dominates has already
been treated above, regardless of the $\widetilde B$ lifetime: In this case,
selectron pair production (p.2 in (\ref{proc})) looks like in the MSSM. Taking
into account the dependence of this branching ratio on $M_{1/2}$, the
corresponding efficiencies and numbers of background/observed events, one finds
that the region (\ref{liminv}) can be completely excluded. (As a matter of
fact, since the decay channel c) plays no role for displaced vertices, the bino
decays always invisibly in this region of the parameter space.) Therefore, the
remaining decay channels for a bino with $M_{\widetilde B} \gsim 43$~GeV are:
b) $\widetilde B \to \widetilde S l^+ l^-$ and d) $\widetilde B \to \widetilde
S \gamma$. In the situation of a macroscopic length of flight, the cases of a
$\widetilde B$ decay inside or outside the detector have to be treated
separately. 

If $\widetilde B$ decays inside the detector ('mesoscopic' decay length: 1~cm$
\lsim l_{\widetilde B} \lsim $3~m, where $l_{\widetilde B}$ denotes the decay
length  in the lab. system), the following topologies are possible:

$\bullet$ The processes p.2 (charged slepton pair production) and p.4 (chargino
pair production) give rise to 2 acoplanar leptons from the primary vertex plus
neutral displaced clusters (lepton pairs or photons) due to delayed $\widetilde
B$ decay. Searches for events with neutral clusters have not been published up
to now, due to vetos against such clusters in order to remove the background
from radiative events \cite{2eA,2eD,2eL,2eO}. However, for small values of
$\Delta M$ (mainly when $\widetilde B$ decays dominantly into $\widetilde S
\gamma$) such neutral clusters could be soft enough not to be vetoed (cf. the
discussion above on photons in the final state). In this case, the limit on the
selectron mass in the MSSM leads to the lower limit (\ref{limsel}) on $M_{1/2}$
($M_{\widetilde B}$). 

$\bullet$ The process p.1 (bino pair production) leads to events with just
neutral displaced vertices and no activity at the primary vertex. Since, in
this case, $\widetilde B$ is the lightest visible particle of the model, this
process would allow to test a larger region in the parameter space than the
processes p.2 and p.4 discussed above. The expected event rates are as in
Fig.~\ref{fig1} for $M_{1/2} \gsim 150$~GeV. If $\Delta M$ is not too small,
the decay product of $\widetilde B$ would be charged leptons with at least 33\%
taus. Clearly this topology is the most difficult one to detect (since triggers
around the primary vertex will not be active\footnote{One could use, however,
an initial state radiative photon to trigger the event.}), and no constraints
on such processes have been published. On the other hand, for small $\Delta M$,
the photonic decay channel d) dominates. Searches have been performed within
the MSSM for $\chi_2^0$ pair production followed by a delayed $\chi_2^0 \to
\chi_1^0 \gamma$ decay \cite{2gD}. However, the efficiency for small mass
differences is tiny and this channel cannot be used. In the region of the
parameter space where this decay channel dominates, the relevant topology is 2
acoplanar electrons arising from selectron pair production, the photons being
soft enough for being accepted as extra neutral clusters in this search (cf.
above). 

If $\widetilde B$ decays outside the detector ('macroscopic' decay length:
$l_{\widetilde B} >$3~m), the situation in the (M+1)SSM with a singlino LSP is
clearly the same as in the corresponding MSSM with $\widetilde B$ being the
true LSP. In particular, the MSSM constraint on the selectron mass can be
applied directly with the additional benefit that $m_{\widetilde e_R} -
M_{\widetilde B}$ is known in terms of $m_{\widetilde e_R}$. Hence, the lower
limit on $M_{1/2}$ ($M_{\widetilde B}$) is given by (\ref{limsel}). 

The present constraints for the various ranges of $M_{1/2}$ (or $M_{\widetilde
B}$) and the various $\widetilde B$ lifetimes can be summarized in
Fig.~\ref{fig2}. On the bottom horizontal line of Fig.~\ref{fig2}, we plot
$M_{1/2}$ in the range of interest, and on the top horizontal line we indicate
the corresponding values of $M_{\widetilde B}$ (with $\lsim$5~GeV accuracy). On
the vertical axis, we plot the $\widetilde B$ decay length in the laboratory
system. In this plane we have indicated in grey those regions (for to
$l_{\widetilde B} >$1~cm), which are excluded by negative results from
acoplanar electron searches. For $l_{\widetilde B} <$1~cm the total number of
events with 4 charged fermions and missing energy in the final state exceeds 20
in the striped region. 

As mentioned above, in the (M+1)SSM with a singlino LSP, the NLSP could
possibly be a stau. Then, limits from MSSM stau searches can be applied. 
Again, if $\lambda$ (or $m_{\widetilde\tau_1} - M_{\widetilde S}$) is
sufficiently small, the $\widetilde\tau_1$ lifetime can become large and give
rise to displaced vertices. Medium or long-lived charged scalars have been
searched for at LEP2 \cite{2eA,longL,GMSBstau}, and the corresponding
constraints can also be applied here. However, the lower limit on stau masses
does not correspond to a definite region in the $(l_{\widetilde B},M_{1/2})$
plane of Fig.~\ref{fig2} which is or not excluded, since even for large values
of $M_{1/2}$, $m_{\widetilde\tau_1}$ can still be relatively small. (Of course,
$\widetilde B$ pair production can still be used, where now the $\widetilde B$
decays always through the cascade $\widetilde B \to \widetilde\tau_1 \tau \to
\widetilde S \tau \tau$. Hence, the $\widetilde B$ lifetime is always very
small. If, in addition, the stau lifetime is also small, processes p.1 and p.3
in (\ref{proc}) give rise to the same topology as in the case of a bino NLSP: 4
charged leptons (taus) plus missing energy. As discussed before, this case is
included in Fig.~\ref{fig1}.)

\section{Summary and outlook}

We have seen that the final state topologies of the (M+1)SSM with a singlino
LSP can differ considerably from the MSSM, due to the additional $\widetilde B
\to \widetilde S X$ cascade. Since these topologies can be the first sign of
sparticle production at LEP2, it is very important to identify and to look
carefully for them. 

In the present paper we have identified these topologies, and studied the
parameter space of the model in order to check whether there are regions not
excluded by negative results from MSSM-like sparticles searches, though
accessible at LEP2 (i.e. with a reasonable expected number of events). 

Indeed we found several such regions, and the associate topologies have been
listed in Table 1: First, we can have 4 charged fermions of various kinds and
missing energy in the final state. Such final states have been looked for in
the context of the MSSM, e.g. in stop and neutralino searches, or in models
with R parity violation. However, the corresponding efficiencies within the
present model are not known up to now. 

In Fig.~\ref{fig1} we have shown the total number of events which can be
expected within the present model as a function of $M_{1/2}$ (which can be
translated into $M_{\widetilde B}$ using (\ref{mbino})). Clearly, assuming a
small but non vanishing efficiency for the topologies of the present model, the
region $M_{1/2} \lsim 140$~GeV, corresponding to $> O(10^2)$ total events,
could already be excluded from searches for 4 charged fermions. Of particular
interest is, however, the region $M_{1/2} \gsim 150$~GeV where only $\widetilde
B$ pair production contributes to this topology; this process allows to test
the largest region in the parameter space. With the corresponding efficiencies
at hand one could expect, e.g., a sensitivity to a total number of $N > 20$ of
4 charged fermion events plus missing energy, which would allow to test the
region up to $M_{1/2} \lsim 190$~GeV (or $M_{\widetilde B} \lsim 75$~GeV) as
indicated by the horizontal line in Fig.~\ref{fig1}, or the striped region in
Fig.~\ref{fig2}. Note that final states with 6 charged fermions can only appear
after slepton or chargino pair production (processes p.2 and p.4 in
(\ref{proc})). The accessible parameter space is thus smaller than the one
covered by $\widetilde B$ pair production.

If the decay length of $\widetilde B$ is mesoscopic (1~cm$\lsim l_{\widetilde
B} \lsim$3~m) and $\widetilde B$ decays visibly, new topologies appear: Either
two leptons at the primary vertex (from slepton or chargino pair production)
plus neutral displaced clusters due to the delayed $\widetilde B$ decay, or
just neutral displaced clusters from $\widetilde B$ pair production. The latter
process is even more promising since it allows to test a larger region in the
parameter space, although it is certainly the most difficult to trigger on. 

Again, the total number of expected events, as a function of $M_{1/2}$ (or
$M_{\widetilde B}$), can be deduced from Fig.~\ref{fig1}. Now, however, the
estimation of the corresponding efficiencies is much more delicate. On the
other hand, the decay channel c) $\widetilde B \to \widetilde S S$ never
appears in this range of the decay length $l_{\widetilde B}$, and the number of
possible final states is reduced. (Now, the region $M_{1/2} \lsim 125$~GeV can
already be excluded: Here the bino decays nearly always invisibly, and the
negative results from acoplanar leptons plus missing energy searches --
associated with the process p.2 in (\ref{proc}) -- can be applied. This is
indicated in Fig.~\ref{fig2} in form of the grey region for 1~cm$\lsim
l_{\widetilde B} \lsim$3~m.)

Herewith we would like to encourage searches for these unconventional
topologies, in order to cover the entire parameter space of the (M+1)SSM with a
singlino LSP. If no excesses are observed at LEP2, we will have to turn to
larger c.m. energies at the Tevatron (Run II), the LHC or -- hopefully -- the
NLC. Again, the (M+1)SSM with a singlino LSP predicts unconventional signals
for these machines, like additional decay cascades (as compared to the MSSM) or
displaced vertices. The details of these topologies and the expected event
rates as a function of the parameters of the (M+1)SSM will have to be
considered in the near future.

\vspace{1cm}
\noindent{\Large\bf Acknowledgments}
\vspace{.5cm}

It is a pleasure to thank L. Duflot for helpful comments. Many useful
discussions in the framework of the French workshop ``GDR Supersym\'etrie'' are
also acknowledged.

\newpage
\section*{Table Caption}
\newcounter{tab}
\begin{list}{\bf Table \arabic{tab}:}{\usecounter{tab}}
\item 
Visible final states after sparticle production in the case of 
microscopic
vertices. (The $\widetilde{B}$ decay $\widetilde{B} \to \widetilde{S}S$ does 
not appear in the case of displaced vertices, see sect. 3.4.) The leptons 
$\ell^+\ell^-$ can be leptons of any generation, including taus (which
are possibly dominant). We have omitted photons from the decay 
$\widetilde{B} \to
\widetilde{S}\gamma$, since these are always soft, see sect. 3.3.  
\label{tab1}
\end{list}

\section*{Figure Captions}

\newcounter{fig}
\begin{list}{\bf Figure \arabic{fig}:}{\usecounter{fig}}

\item Number of 4 charged fermion events expected from $\widetilde B$,
$\widetilde e_R$, $\widetilde\tau_1$, $\chi_1^\pm$ pair production as a
function of $M_{1/2}$. \label{fig1} 

\item Regions in the plane $l_{\widetilde B}$ (in the laboratory system) vs. 
$M_{1/2}$, which are excluded due to negative results from searches for
acoplanar electrons at LEP2, are indicated in grey. (On the top horizontal line
we indicate the corresponding values of $M_{\widetilde B}$, with $\lsim$5~GeV
accuracy). The remaining regions still have to be explored. In the striped
region, for $l_{\widetilde B} <$ 1~cm, the total number of events with 4
charged fermions and missing energy in the final state exceeds 20. The vertical
dashed line indicates the kinematic limit for $\widetilde B$ pair production at
LEP2 with $\sqrt{s} = 183$~GeV. \label{fig2}

\end{list}

\textheight 250truemm

\newpage
\pagestyle{empty}
\voffset=-1 truecm
\hoffset=-1 truecm

\begin{tabular} {|l|l|l|l|l|} 
\hline
&\multicolumn{4}{c|}{Production Process} \\
\cline{2-5}
& & & & \\
&\qquad p.1 &\qquad p.2 &\qquad p.3 &\qquad p.4 \\
&$e^+e^- \to \widetilde{B}_1 \widetilde{B}_2$
&$e^+e^- \to \widetilde{l}_R \widetilde{l}_R^*$
&$e^+e^- \to \widetilde{\nu} \widetilde{\nu}^*$
&$e^+e^- \to \chi_1^+ \chi_1^-$ \\
& &$\to l^+ \widetilde{B}_1 l^- \widetilde{B}_2$
&$\to \nu \widetilde{B}_1 \bar{\nu}\widetilde{B}_2$
&$\to l^+\widetilde{\nu} l '^{-} \widetilde{\nu}^*$ \\
& & & &$\to l^+\nu \widetilde{B}_1 l'^{-} \bar{\nu} \widetilde{B}_2$ \\
\cline{0-0}
\qquad Bino Decays & & & & \\ 
\hline
& & & & \\
$\widetilde{B}_1 \to \widetilde{S} \nu \bar{\nu}$ &0
&$l^+l^- + {/ \hskip - 3 truemm E_T}$ &0
&$l^+l'^- + {/ \hskip - 3 truemm E_T}$ \\
$\widetilde{B}_2 \to \widetilde{S} \nu \bar{\nu}$ & & & & \\
&\qquad \qquad (1.1) &\qquad \qquad (1.2) &\qquad \qquad (1.3)
&\qquad \qquad (1.4) \\

\hline
& & & & \\
$\widetilde{B}_1 \to \widetilde{S} \nu \bar{\nu}$ 
&$l^+l^- + {/ \hskip - 3 truemm E_T}$
&$l^+l^- l^+l^- + {/ \hskip - 3 truemm E_T}$
&$l^+l^- + {/ \hskip - 3 truemm E_T}$
&$l^+l^-l^+l '^- + {/ \hskip - 3 truemm E_T}$ \\
$\widetilde{B}_2 \to \widetilde{S} l^+ l^-$ & & & & \\ 
&\qquad \qquad (2.1) &\qquad \qquad (2.2) &\qquad \qquad (2.3) 
&\qquad \qquad (2.4) \\

\hline
& & & & \\
$\widetilde{B}_1 \to \widetilde{S} \nu \bar{\nu}$
&$b\bar{b} + {/ \hskip - 3 truemm E_T}$
&$l^+l^-b \bar{b} + {/ \hskip - 3 truemm E_T}$
&$b\bar{b} + {/ \hskip - 3 truemm E_T}$
&$l^+l'^-b\bar{b} + {/ \hskip - 3 truemm E_T}$ \\
$\widetilde{B}_2 \to \widetilde{S}S \to \widetilde{S} b \bar{b}$ & & & & \\
&\qquad \qquad (3.1) &\qquad \qquad (3.2) &\qquad \qquad (3.3)
&\qquad \qquad (3.4) \\

\hline
& & & & \\
$\widetilde{B}_1 \to \widetilde{S} \nu \bar{\nu}$
&$\tau^+\tau^- + {/ \hskip - 3 truemm E_T}$
&$l^+l^-\tau^+\tau^- + {/ \hskip - 3 truemm E_T}$
&$\tau^+ \tau^- + {/ \hskip - 3 truemm E_T}$
&$l^+l'^-\tau^+\tau^- + {/ \hskip - 3 truemm E_T}$ \\
$\widetilde{B}_2 \to \widetilde{S}S \to \widetilde{S} \tau^+ \tau^-$ & & & & \\
&\qquad \qquad (4.1) &\qquad \qquad (4.2) &\qquad \qquad (4.3)
&\qquad \qquad (4.4) \\

\hline
& & & & \\
$\widetilde{B}_1 \to \widetilde{S} l^+ l^-$
&$l^+l^-l^+l^- + {/ \hskip - 3 truemm E_T}$
&$l^+l^-l^+l^-l^+l^-$
&$l^+l^-l^+l^- + {/ \hskip - 3 truemm E_T}$
&$l^+l^-l^+l^-l^+l '^-$ \\
$\widetilde{B}_2 \to \widetilde{S} l^+ l^-$
& &$+ {/ \hskip - 3 truemm E_T}$ & &$+ {/ \hskip - 3 truemm E_T}$ \\
&\qquad \qquad (5.1) &\qquad \qquad (5.2) &\qquad \qquad (5.3)
&\qquad \qquad (5.4) \\

\hline
& & & & \\
$\widetilde{B}_1 \to \widetilde{S} l^+ l^-$
&$l^+l^-b \bar{b} + {/ \hskip - 3 truemm E_T}$
&$l^+l^-l^+l^-b\bar{b} + {/ \hskip - 3 truemm E_T}$
&$l^+l^-b\bar{b} + {/ \hskip - 3 truemm E_T}$
&$l^+l^-b\bar{b} l^+l '^- + {/ \hskip - 3 truemm E_T}$ \\
$\widetilde{B}_2 \to \widetilde{S}S \to \widetilde{S}b\bar{b}$ & & & & \\
&\qquad \qquad (6.1) &\qquad \qquad (6.2) &\qquad \qquad (6.3)
&\qquad \qquad (6.4) \\

\hline
& & & & \\
$\widetilde{B}_1 \to \widetilde{S} l^+ l^-$
&$l^+l^-\tau^+ \tau^- + {/ \hskip - 3 truemm E_T}$
&$l^+l^-l^+l^-\tau^+\tau^-$
&$l^+l^-\tau^+\tau^- + {/ \hskip - 3 truemm E_T}$
&$l^+l^-\tau^+\tau^- l^+l '^-$ \\
$\widetilde{B}_2 \to \widetilde{S}S \to \widetilde{S}\tau^+\tau^-$
& &$+ {/ \hskip - 3 truemm E_T}$ & &$+ {/ \hskip - 3 truemm E_T}$ \\
&\qquad \qquad (7.1) &\qquad \qquad (7.2) &\qquad \qquad (7.3)
&\qquad \qquad (7.4) \\

\hline 
& & & & \\
$\widetilde{B}_1 \to \widetilde{S}S \to \widetilde{S} b\bar{b}$
&$b\bar{b}b\bar{b} + {/ \hskip - 3 truemm E_T}$
&$l^+l^-b\bar{b}b\bar{b} + {/ \hskip - 3 truemm E_T}$
&$b\bar{b}b\bar{b} + {/ \hskip - 3 truemm E_T}$
&$b\bar{b}b\bar{b} l^+l '^- + {/ \hskip - 3 truemm E_T}$ \\
$\widetilde{B}_2 \to \widetilde{S}S \to \widetilde{S}b\bar{b}$ & & & & \\
&\qquad \qquad (8.1) &\qquad \qquad (8.2) &\qquad \qquad (8.3)
&\qquad \qquad (8.4)\\

\hline 
& & & & \\
$\widetilde{B}_1 \to \widetilde{S}S \to \widetilde{S} \tau^+\tau^-$
&$\tau^+\tau^-\tau^+\tau^- + {/ \hskip - 3 truemm E_T}$
&$l^+l^-\tau^+\tau^-\tau^+\tau^-$
&$\tau^+\tau^-\tau^+\tau^- + {/ \hskip - 3 truemm E_T}$
&$\tau^+\tau^-\tau^+\tau^- l^+l '^-$ \\
$\widetilde{B}_2 \to \widetilde{S}S \to \widetilde{S}\tau^+\tau^-$
& &$+ {/ \hskip - 3 truemm E_T}$ & &$+ {/ \hskip - 3 truemm E_T}$ \\
&\qquad \qquad (9.1) &\qquad \qquad (9.2) &\qquad \qquad (9.3)
&\qquad \qquad (9.4)\\

\hline 
\end{tabular}

\vskip 15 truemm
\hskip 80 truemm {\bf Table 1}

\newpage

\hoffset=1truecm

\begin{figure}[p]
\unitlength1cm
\begin{picture}(12,20)
\put(6.5,0){\bf Figure 1}
\put(-1.5,11){$\bf N(4f + {/ \hskip - 3 truemm E_T})$}
\put(7,2.5){$\bf M_{1/2} [GeV]$}
\put(0,0){\epsffile{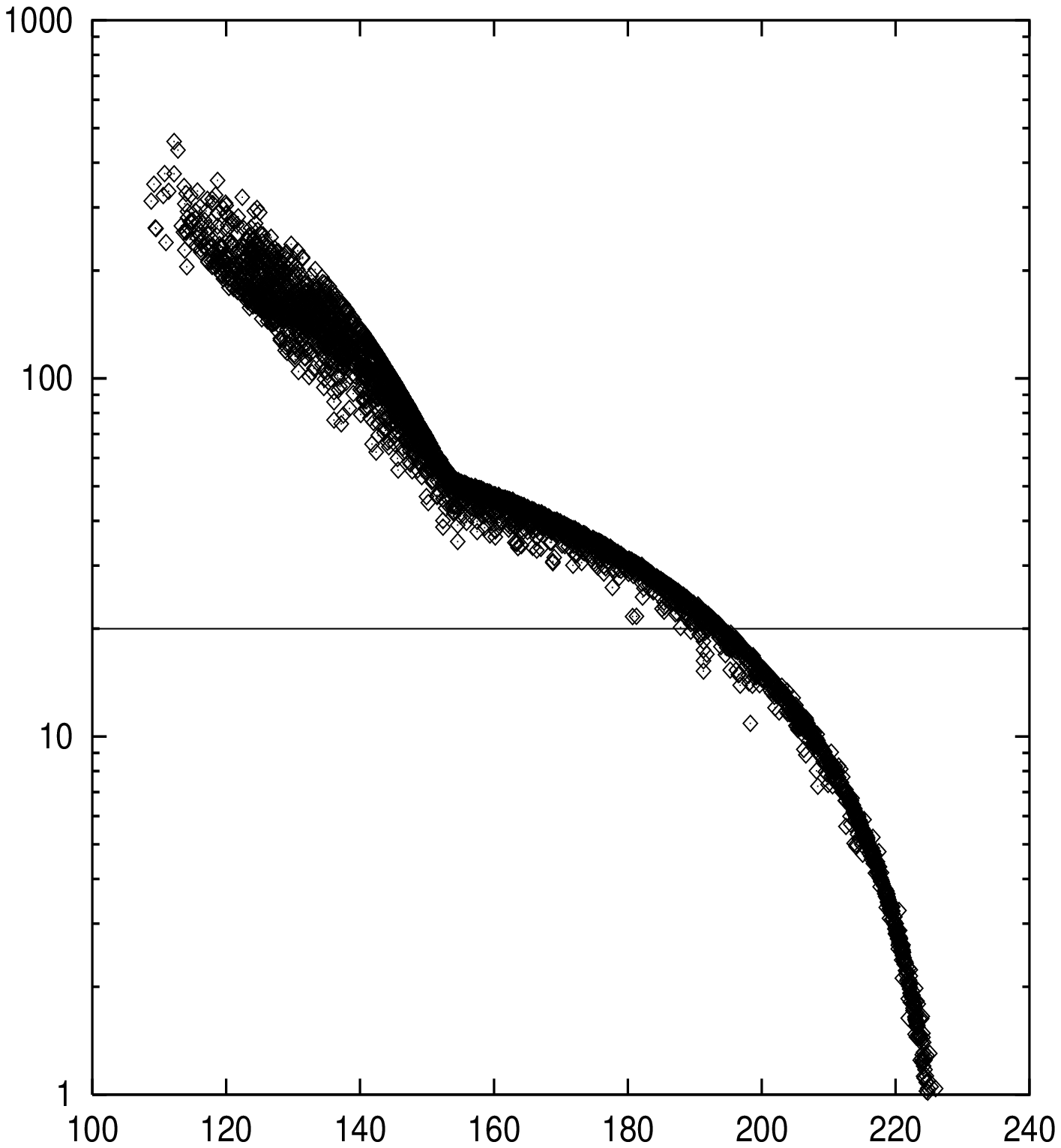}}
\end{picture}
\end{figure}

\begin{figure}[p]
\unitlength1cm
\begin{picture}(12,20)
\put(6.5,0){\bf Figure 2}
\put(-1,11){$\bf l_{\widetilde B} [m]$}
\put(7,2.5){$\bf M_{1/2} [GeV]$}
\put(7,20){$\bf M_{\widetilde B} [GeV]$}
\put(0,0){\epsffile{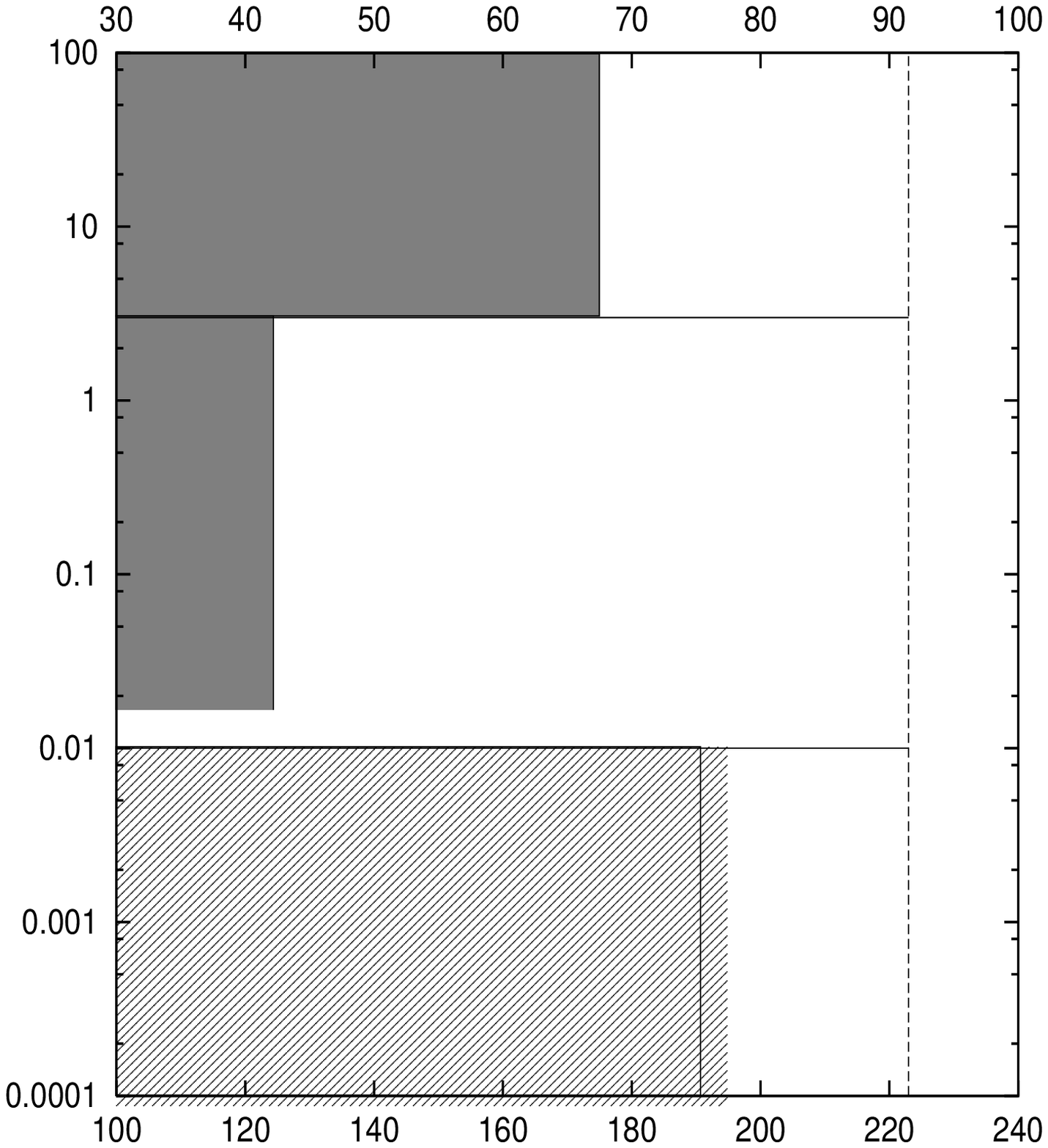}}
\end{picture}
\end{figure}

\end{document}